\documentclass[11pt]{article} 

\usepackage[titlenumbered,ruled]{algorithm2e} 
\usepackage[margin=1.2in]{geometry}
\usepackage{natbib} 
\usepackage{graphicx} 
\usepackage{booktabs}
\usepackage{url} 
\usepackage{subfig} 

\usepackage{amsmath,amsthm,amssymb} 
\usepackage{bbold} 
\usepackage{authblk}

\theoremstyle{definition}

\usepackage{color}

\usepackage[font={small,it}]{caption}

\begin{document}

\title{{\Large Continuously Updated Data Analysis Systems}}
\author[1]{Lee F. Richardson}
\affil[1]{Department of Statistics and Data Science, Carnegie Mellon University}
\date{\today}
\maketitle

\begin{abstract}
When doing data science, it's important to know what you're building. This paper describes an idealized final product of a data science project, called a Continuously Updated Data-Analysis System (CUDAS). The CUDAS concept synthesizes ideas from a range of successful data science projects, such as Nate Silver's {\it FiveThirtyEight}. A CUDAS can be built for any context, such as the state of the economy, the state of the climate, and so on. To demonstrate, we build two CUDAS systems. The first provides continuously-updated ratings for soccer players, based on the newly developed {\it Augmented Adjusted Plus-Minus} statistic. The second creates a large dataset of {\it synthetic ecosystems}, which is used for agent-based modeling of infectious diseases.
\end{abstract}

\section{Introduction}
\label{sec:intro}
When I work on data science projects, it helps to imagine what the final product will look like. At the end of the rainbow, what's my pot of gold? In most projects, the final product is defined for us; your boss wants a report, your software engineer friend wants a script, your advisor wants a paper, and so on. But we'll forget these constraints here, and instead describe an idealized final product for a data science project, called a Continuously-Updated Data-Analysis System (CUDAS).

The CUDAS concept isn't new, and you've probably seen it before. For example, an incredibly popular CUDAS was the 2016 election forecast by {\it FiveThirtyEight}. What makes this a CUDAS? I'll explain next. 

\section{The CUDAS Concept}
\label{sec:cudas}
Broadly speaking, a CUDAS has three components:

\begin{enumerate}
	\item {\bf Data Pipeline}
	\item {\bf Data Analysis}
	\item {\bf Continuously-Updated Results}
\end{enumerate}

The data pipeline processes the raw data so it's ready for analysis, the data-analysis converts the processed data into the results we're interested in, and we continuously update our results as new data comes in. These components don't need to happen in a sequence. For example, we may need to update our data pipeline after realizing that our data-analysis is missing a covariate. 

These three components closely follow definitions of data science described elsewhere (e.g. \cite{silver2014fox}, \cite{donoho201750}, and \cite{wickham2016r}). For example, I adapted the definition given in Chapter 1 of \cite{wickham2016r} by grouping the {\it import} and {\it tidy} boxes into a single component, called the data-pipeline. I've also grouped {\it transform}, {\it visualize}, and {\it model} into a single component, called {\it data-analysis} (\cite{tukey1962future}). Finally, I changed {\it communicate} to {\it continuously-updated results}, because a CUDAS updates when new data becomes available.

This CUDAS definition also closely follows the definition of {\it Greater Data Science} given by \cite{donoho201750}. In fact, I think about a CUDAS as an implementation of this framework, since each of Donoho's six divisions, even {\it science about data science}, can be viewed in terms of their impact on a CUDAS. More on this later. 

\subsection{Three example CUDAS projects}
\label{sec:examplecudas}
I got the idea for a CUDAS by studying successful data science projects, and trying to abstract what they had in common. I'll walk through my three favorite examples next. 

\subsubsection{FiveThirtyEight's 2016 Election Forecast}
\label{sec:fivethirtyeight}
\begin{figure}[!ht]
    \centering
    \subfloat{{\includegraphics[width=6cm]{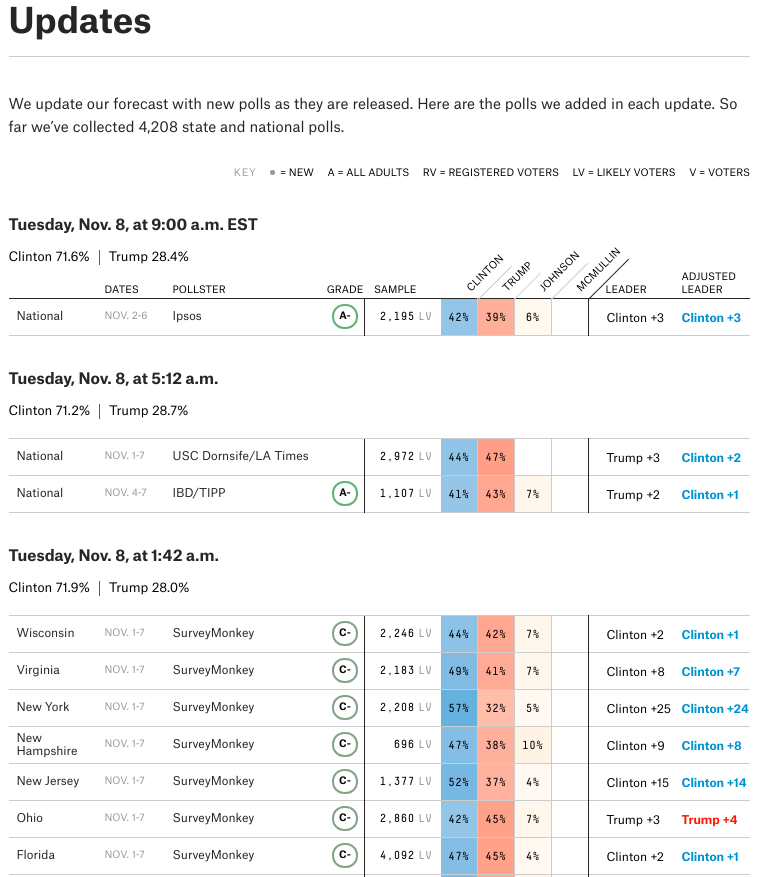} }}
    \qquad
    \subfloat{{\includegraphics[width=6cm]{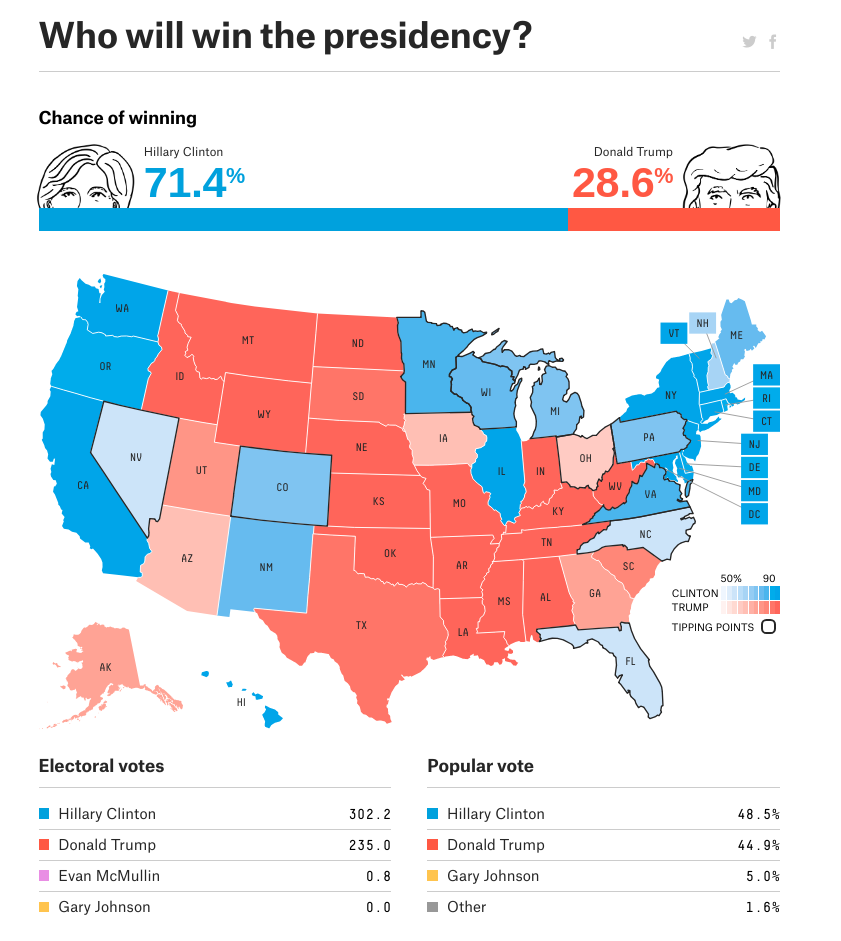} }}
    \caption{{\bf Left}: A visualization of the polls that are collected, adjusted, then added to the forecast. {\bf Right}: The predicted probability of each candidate winning the 2016 presidential election.}
    \label{fig:fivethirtyeight}
\end{figure}

The 2016 election forecast from {\it FiveThiryEight} was (perhaps) the most popular CUDAS of all time. Their system collects polling data, uses this data to forecast the probability of each candidate winning, and continuously-updates the forecast on a beautiful interactive web page. Figure \ref{fig:fivethirtyeight} shows two key screen shots of the project. 

\subsubsection{The Global Burden of Disease}
\label{sec:gbd}
Another great CUDAS comes from the {\it Global Burden of Disease} (GBD) study, produced by the {\it Institute of Health Metrics and Evaluation} (\cite{lopez1998global}). The GBD is an extremely ambitious project, with a goal of collecting and synthesizing all the world's health data, and providing continuously-updated estimates of disease burden. The GBD is a scientific triumph, and the book {\it Epic Measures} by \cite{smith2015epic} chronicles the story from its beginnings. 
	
\begin{figure}[!ht]
\centering
\includegraphics[width=\textwidth]{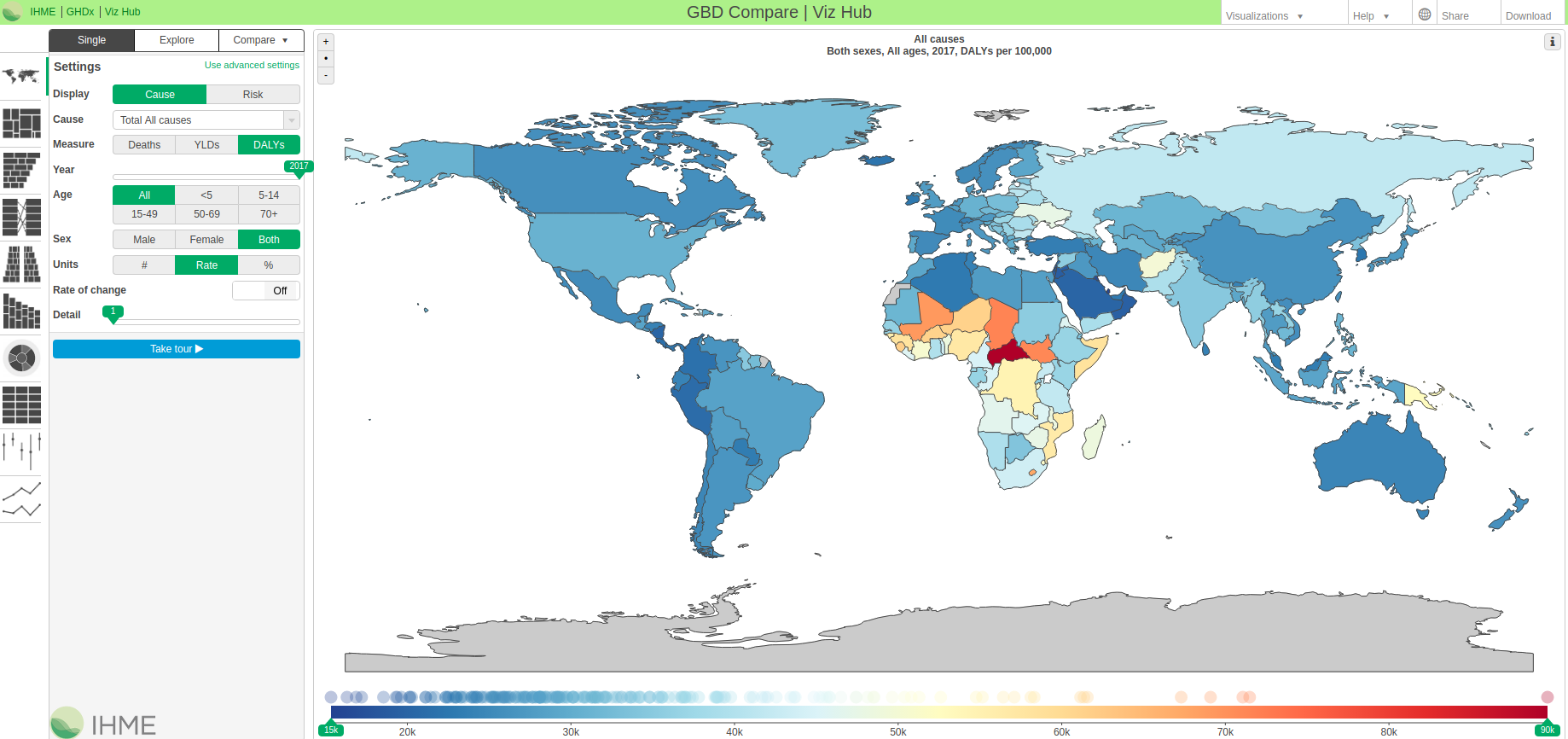}
\caption{The GBD's interactive website that displays the DALY's attributed to all diseases, by location, in 2017. The GBD produces multiple interactive visualizations, updates annually, and given users access to a wide variety of results.}
\label{fig:gbd}
\end{figure}

Let's think about the GBD in terms of a CUDAS. First, the GBD employs a team whose goal is collecting all the health data they can get their hands on, from surveys, to scientific literature, to vital registration systems, and more. Next, the GBD has a team of disease experts, statisticians, computer scientists, epidemiologists, etc. to model the burden of each individual disease. Then, the individual disease estimates are combined into a single metric, called the Disability-Adjusted Life Year (DALY, \cite{murray1997understanding}). Finally, the GBD provides spectacular interactive visualizations of their results, which they update annually, an example of which is shown in Figure \ref{fig:gbd}.

\subsubsection{WAR on Ice}
Our final CUDAS example is the {\tt war-on-ice} project, available online at:

\begin{center}
	\url{http://war-on-ice.com/}.
\end{center}

Although the project is no longer active\footnote{The two creators are now employed by professional hockey teams.}, at it's apex, {\tt war-on-ice} provided advanced hockey statistics that updated after every night of games. Notably, their statistics and visualizations were fairly advanced (e.g. \cite{thomas2013competing}), compared with statistics available on other websites.

An interesting twist on the {\tt war-on-ice} CUDAS is that the author's made an critical piece of their data pipeline, the {\tt nhlscrapr R} package, available. In later years, one of the author's has been behind a similar {\tt nflscrapR} package for American football (\cite{horowitz2017nflscrapr}), which shows the early signs of a generalizable idea.

The 2016 election forecast, the GBD, and {\tt war-on-ice} come from completely different contexts (politics, global health, and hockey). But when viewed from a CUDAS lens, the projects are similar. The next section provides more detail on the similarities between these three systems.

\subsection{Elements of a CUDAS}
\label{sec:cudaselements}
What do the 2016 Election forecast, the GBD, and {\tt war-on-ice} have in common? For starters, each project has a data pipeline, data-analysis, and continuously updated results. But each project also understood the dependencies between these components: the data-analysis is nothing without the data pipeline, and the data-analysis isn't as valuable without the continuously-updated results. Let's go into more detail on what made these three CUDAS systems stand out.

{\bf Multiple data sources are synthesized in a purposeful way}.
In each of our three examples, the data was {\it available} online, but the data wasn't formatted for data-analysis. For example, the 2016 election forecast collects polls from many different sources, the GBD combines different data types from many different diseases, and {\tt war-on-ice} collects play-by-play data, images, box score statistics, and more. 

But these projects didn't just {\it collect} the data, they also knew what to do with it. The data was rigorously extracted and transformed into the precise format required for the data-analysis. Nate Silver provides a detailed user guide to the 2016 election forecast (\cite{electionguide2016}), in which he describes the critical steps of adjusting the polls, and combining the polls with other data sources, such as economic data. So, it's not enough for a data pipeline to collect the data, a good pipeline must also know how the raw data needs to be processed in order to produce the results the CUDAS is ultimately interested in. 

{\bf The results are interesting and interpretable}.
The data-analysis performed in our three examples isn't extremely complicated, but it's not trivial either. Each of our examples uses some sort of statistical model: \cite{election2016} uses a Bayesian approach to forecast who will win the election, the GBD uses many models (e.g. {\it DisMod} (\cite{dismod2019}), and {\tt war-on-ice} implements the adjusted plus-minus methodology described in \cite{thomas2013competing}.

In my view, these models are successful because they're {\it interesting} and {\it interpretable}. By interesting, I mean that each model was able to gain a large following of users who wanted to know how the results changed as new data came in. By interpretable, I mean that the output of the model was easy to understand: \cite{election2016} gives each candidate a probability of winning, the GBD summarizes disease burden into a single metric (the DALY), and {\tt war-on-ice} ranks players based on their contribution to winning. In each case, it doesn't take rocket scientist to understand the results. 

{\bf The results are continuously updated in a highly intuitive display}.
Finally, and I think most important for their success, each example {\it continuously-updates} their results. And they don't {\it just} update their results, they {\it display} their results in highly intuitive web applications, which gives users a simple way to stay up to date. If there's one thing we've learned in the information age, it's that people like checking their devices for updates (think: Facebook notifications). 

So now we know what a CUDAS is, and we've analyzed three examples. Let's use these insights to create some CUDAS systems of our own.

\section{Building CUDAS systems}
I've luckily been involved in developing several CUDAS systems. Full disclosure: I worked on the GBD project (Section \ref{sec:gbd}) for a year. Since then, I've had a larger role in developing two other CUDAS projects: one for infectious disease modeling, and the other for ranking soccer players. In this section, I'll go through the details of building them both. 

\subsection{Synthetic Populations and Ecosystems of the World}
In {\it The Signal and the Noise} (\cite{silver2012signal}), Nate Silver overviews the state of disease forecasting. After discussing the limitations of compartment models, Silver discusses the potential of a new approach, called {\it agent-based models}. For disease modeling, agent-based models simulate the daily interactions of people (e.g. when they talk, when they're in the same room), and track how a disease spreads based on these interactions.\footnote{Silver interviews a group of agent-based modelers from the {\it University of Pittsburgh}, who work on an infectious disease model called {\it FRED: A Framework for Reconstructing Epidemic Dynamics} (\cite{grefenstette2013fred}). In the CUDAS described in this section, we essentially worked to create a synthetic ecosystems for the FRED model.}

For agent-based models to work, they need a dataset with a record for each person in the population. The dataset should also include where each person lives, where they go to school, and other information relevant to disease modeling. Agent-based modelers refer to these datasets as {\it synthetic ecosystems}. 

Synthetic ecosystems are tricky to build: you need data from different sources, you need to integrate these data sources together, you need to make sure the synthetic ecosystem represents the population, and you need a large computer. And because this is a tricky problem, there's a demand in the agent-based modeling community for high quality synthetic ecosystems. That's where we came in. As part of the MIDAS research network, a group of us were tasked with generating synthetic ecosystems. Our goal: build a CUDAS for synthetic ecosystems.

\subsubsection{Data Pipeline}
To create a synthetic ecosystems, we need to know:

\begin{itemize}
	\item {\bf How many people to create}. For example, to create a synthetic ecosystem for Pittsburgh, we need to know how many people live in Pittsburgh.
	\item {\bf Geography}. Continuing the Pittsburgh example, we need to know how the neighborhoods are organized, where the roads are, where the schools are, and so on. 
	\item {\bf The characteristics of the people (age, gender, occupation, etc.)}.
\end{itemize}

\begin{figure}[!ht]
\centering
\includegraphics[width=\textwidth]{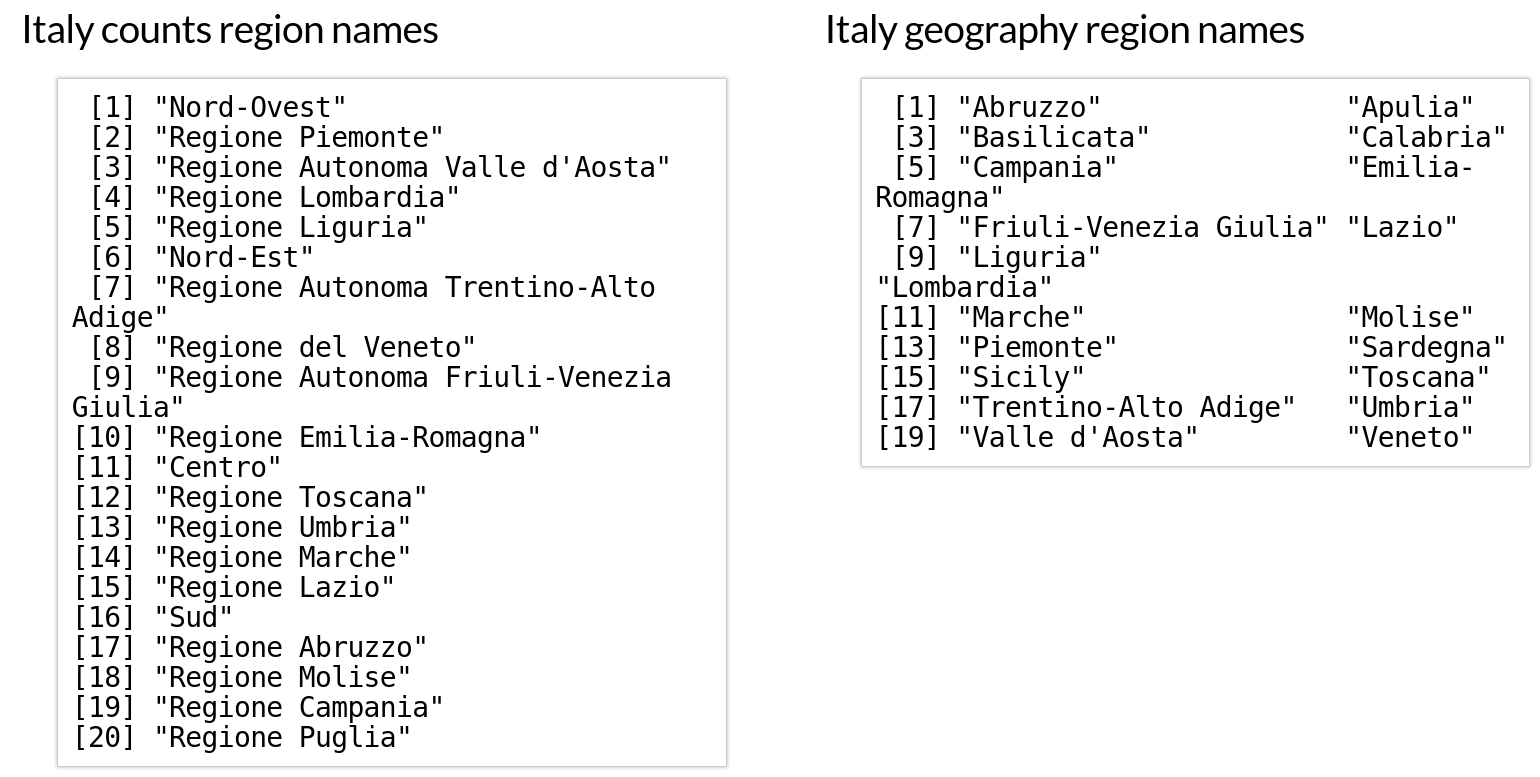}
\caption{A common problem in building data pipeline's for CUDAS systems is matching names across multiple data sources. This figure shows how two different data sources split Italy into 20 smaller regions. Matching names is often the most time consuming part of building a CUDAS, and there is a need for more efficient solutions to this problem.}
\label{fig:harmonization}
\end{figure}

All of this data is available online, but different pieces are available in different locations. So the first part of our data pipeline consisted of scripts to collect data and store them on our computing cluster, hosted by the Pittsburgh Supercomputer Center (\cite{olympus}). Next, we laboriously ensured that each data source shared a common geography. This is difficult, because each data source partitions countries into smaller regions. But unfortunately, each data source {\it differs} in how it partitions countries. For example, the left side of Figure \ref{fig:harmonization} shows how the website {\it GeoHive} (\cite{geohive}) splits Italy into 20 regions, and the right side of Figure \ref{fig:harmonization} shows how IPUMS (\cite{ipumsi}) splits Italy into 20 regions. While these two data sources are {\it close}, it still took a lot of work to make sure that both datasets had the same geographies. And Italy was easy compared with the rest of the countries.  

Thus, a substantial element of our data pipeline involved matching the geographies of different data sources. We did this manually, but in the next section, I'll discuss a more general solution to this problem.

\subsubsection{Data-Analysis}
Once we (finally) had the data ready, we needed a method to turn this data into synthetic ecosystems. For this, we developed the SPEW framework for synthetic ecosystems, which is described in \cite{gallagher2018spew}. The framework samples people, assigns these people to households, schools and workplaces, then assigns locations to the households, schools, and workplaces. We used intuitive algorithms for each of these tasks. For example, we sampled the characteristics of people using {\it microdata}, where microdata is simply a representative sample of the population. And we assigned people to schools based on the location of their household, the location of their school, and the size of each school. 

To implement the SPEW framework, we created the {\tt spew R} package. We used this package to generate all of our synthetic ecosystems, and it was designed to work on the Pittsburgh Supercomputing Cluster, where our data was stored. The package is available online at:

\begin{center}
	\url{https://github.com/leerichardson/spew}.
\end{center}

\subsubsection{Continuously-Updated Results}
After generating the synthetic ecosystems, we needed to make them available to agent-based modelers. First things first, we created a website where users could simply download the complete ecosystems, which is available online at:

\begin{center}
	\url{https://www.epimodels.org/drupal-new/?q=node/160}.
\end{center}

\begin{figure}[!ht]
\centering
\includegraphics[width=\textwidth]{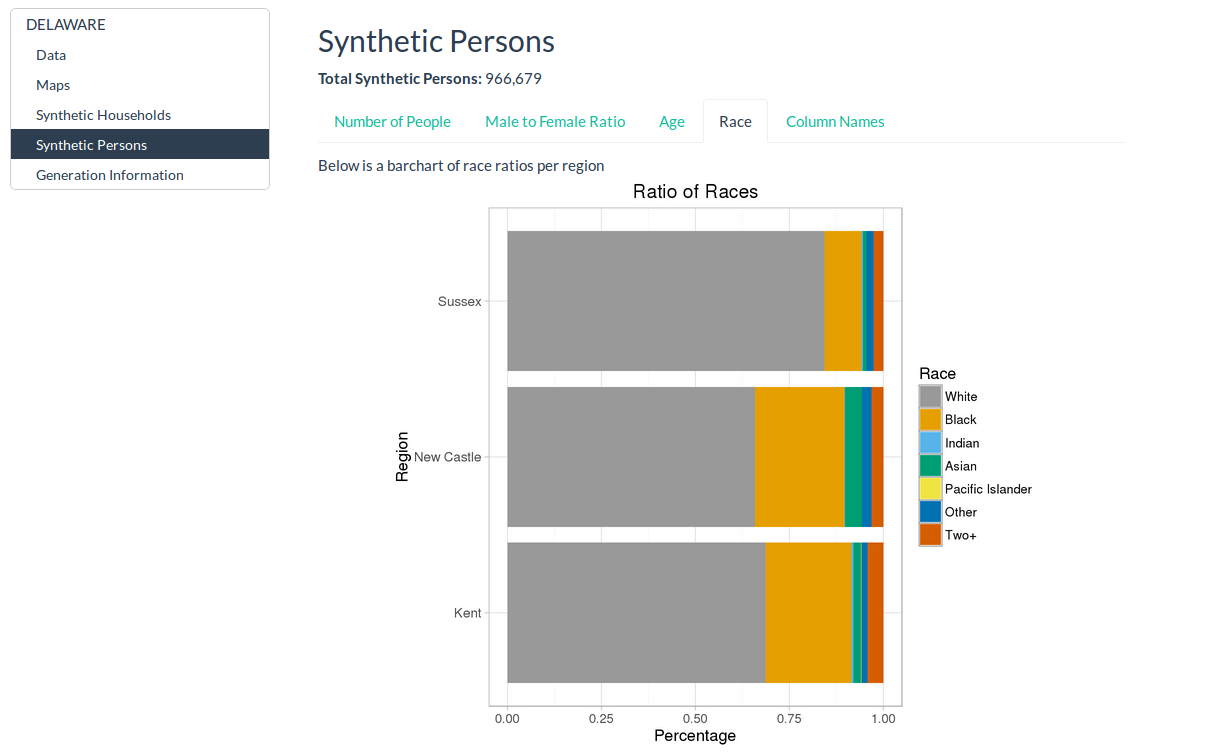}
\caption{Automatically generated reports that summarize each synthetic ecosystem. These reports helped agent-based modelers understand their ecosystems, and they helped us debug our software.}
\label{fig:spew-results}
\end{figure}

But these results weren't very intuitive, so Shannon, a fellow PhD student on this project, wrote a general markdown script that produced summary reports for each synthetic ecosystem (see Figure \ref{fig:spew-results} for an example). Not only did these reports help agent-based modelers understand their synthetic ecosystems, but they also helped us debug our software, and ensured that our synthetic ecosystems passed the {\it intraocular} (``hits you in the eyes'') test.

In terms of continuously-updated results, the idea was that whenever the Census released a new data sample, or {\it GeoHive} released new population counts, or when a new data source became available, a user would be able to pass the new data source through the SPEW framework, and obtain a synthetic ecosystem that accounted for the new data.

\begin{figure}[!ht]
\centering
\includegraphics[height=8cm]{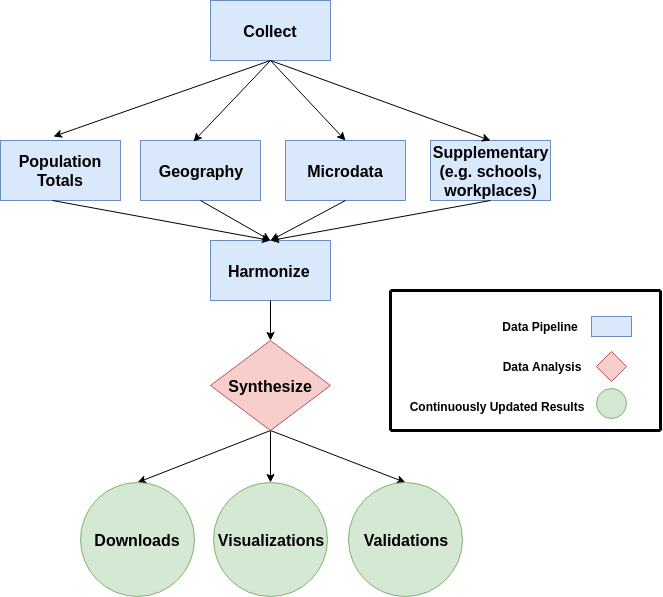}
\caption{The spew framework in from \cite{gallagher2018spew}. From a CUDAS perspective, we clearly see that the framework decomposes into a data pipeline, data-analysis, and continuously-updated results.}
\label{fig:spew-process}
\end{figure}

Although we released several versions of synthetic ecosystems, the newest of which used more recent data, we were never able to reliably and efficiently produce continuously-updated synthetic ecosystems in this idealized manner. But the dream lives on. As a silver lining, we developed a diagram that describes the SPEW framework, shown in Figure \ref{fig:spew-process}. And as the figure shows, this process cleanly decomposes into a CUDAS.

\subsection{A CUDAS for Soccer Ratings}
\label{sec:cudassoccer}
The second CUDAS we'll walk through is for a recently developed soccer metric, called {\it Augmented Adjusted Plus-Minus} (AAPM, \cite{matano2018augmenting}). The details are in the paper, but the basic idea is that AAPM combines two data sources: FIFA ratings and play-by-play data, and uses these data sources to rate each player. In \cite{matano2018augmenting}, it's shown that AAPM predicts game outcomes better than other statistics. 

Why is AAPM a good statistic for a CUDAS? Earlier, I noted that the data-analysis results for a CUDAS should be {\it interesting} and {\it interpretable}. In principle, the AAPM statistic should be interesting to soccer fans, especially since it's tied to predictive accuracy. The AAPM statistic is also interpretable, since it ranks each player, and can be easily displayed in a table.

\subsubsection{Data Pipeline}
\begin{figure}[!ht]
\centering
\includegraphics[width=\textwidth]{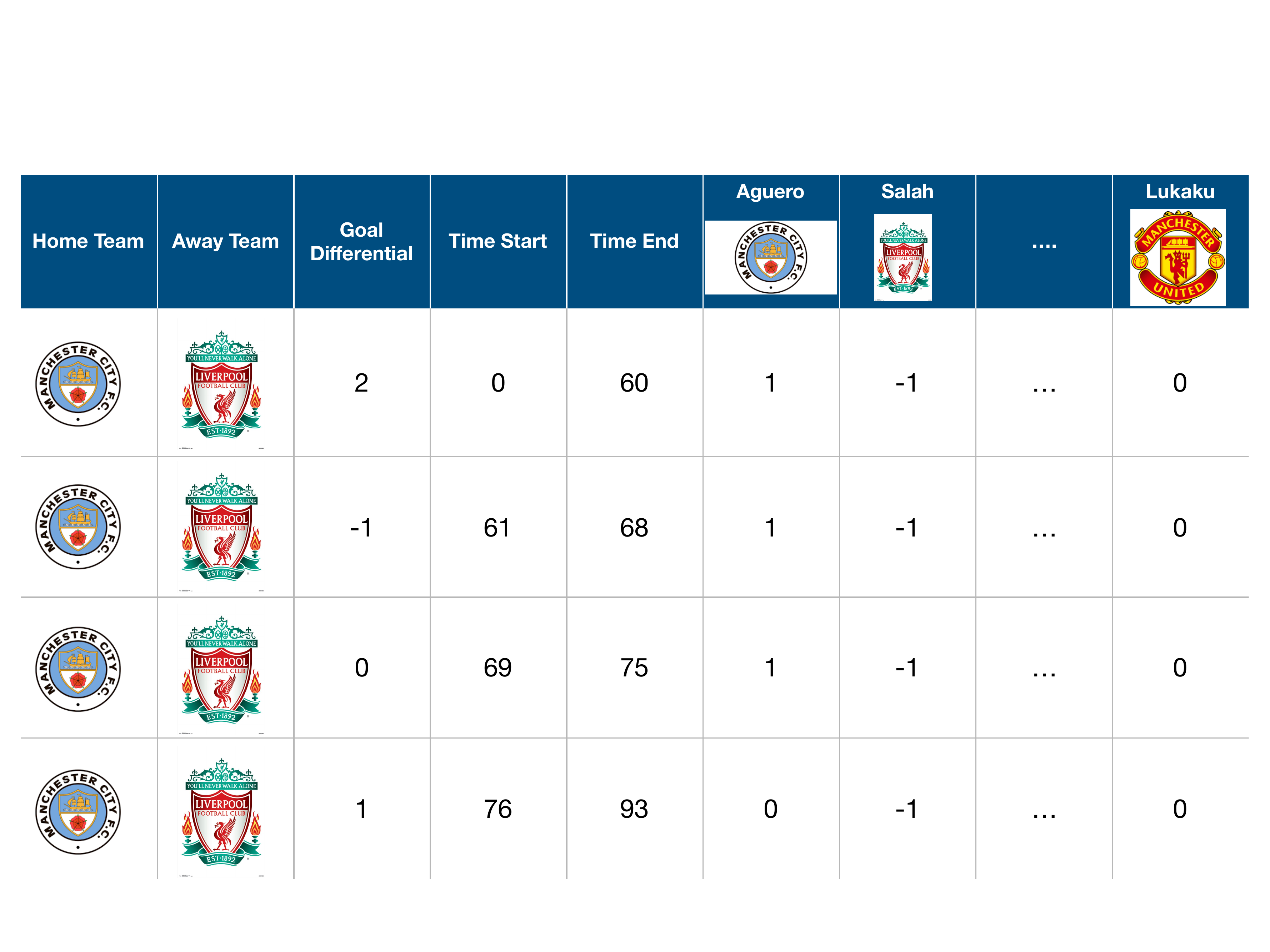}
\caption{The design matrix produced by the data pipeline for our AAPM CUDAS. Each column (the {\bf X} in a regression model) represents a player. This design matrix, and a FIFA rating for each player, is the input to our statistical model that computes the AAPM statistic.}
\label{fig:aapm-design-matrix}
\end{figure}

We need two data sources to compute AAPM: play-by-play information, and FIFA ratings from the beginning of each season. With these two sources, we need to produce a {\it design matrix}, which is the input required for our statistical model that computes AAPM. In the design matrix, each column represents a single player, and Figure \ref{fig:aapm-design-matrix} shows what the design matrix looks like. We also need to link each player (column in the design matrix) with a FIFA rating. 

There are several complications to building the pipeline, such as:

\begin{itemize}
	\item Finding websites with the data, then writing scripts to extract it. 
	\item Matching the names of soccer players from multiple data sources. 
	\item Automating the process so that it works across seasons, leagues, etc.
\end{itemize}

You may have noticed that these challenges are the same we faced when we built our CUDAS for synthetic ecosystems. In each case, we needed to collect data from multiple sources, and match names across each source. For synthetic ecosystems, we matched geographic names, and for soccer ratings, we matched player names.

We overcame these challenges more effectively for the soccer CUDAS. Here, we developed two {\tt R} packages: the first extracts the play-by-play and FIFA data, and the second matches player names using {\it active record linkage}. The data collection package is similar to the {\tt nhlscrapR} package developed by {\tt war-on-ice}, and we actually used a similar package, called {\tt fcscrapR}, for some parts of the collection.

The name matching package, called {\tt arl}, is a bit more interesting. The {\tt arl} package automatically matches the names that are identical in both data sources, partially matches the names that were close, using probabilistic record linkage methods, and manually matches all the remaining names. In short: we automated as much as we could, and manually matched the rest. Given the differences between some data sources, sometimes this is the best you can do. 

Similar to our CUDAS for synthetic ecosystems, the majority of the work was building the data pipeline. For seasoned data scientists, this is an obvious point. 

\subsubsection{Data-Analysis}
Once our data pipeline produced the design matrix with linked FIFA ratings, we were ready to compute AAPM. The AAPM statistic is calculated with a Bayesian regression model, where FIFA ratings are the prior distribution for each player. We developed another {\tt R} package to fit our model:

\begin{center}
	\url{https://github.com/tpospisi/PlusMinusModels},
\end{center}

which relies standard Bayesian software (\cite{carpenter2017stan}). After various model checks and tweaks, we verified that:

\begin{itemize}
	\item Our results passed the {\it intraocular} test (the best/worst players made sense).
	\item Our model predicted game outcomes better than baseline and comparison statistics. 
\end{itemize}

With our results in hand, the final step was producing the continuously-updated results.

\subsubsection{Continuously Updated Results}
To produce continuously updated results, we need to answer two questions:

\begin{enumerate}
	\item What's the best way to display our results?
	\item How can we continuously-update them?
\end{enumerate}

To display the results, we followed another successful CUDAS: ESPN's {\it Real Plus-Minus} statistic (RPM, \cite{espnrpm}). ESPN displays the RPM statistic in a simple table, which users can sort by offense, defense, or position. We created a similar table, which is available online at:

\begin{center}
	\url{https://www.intraocular.net/cudas/aug-apm/}.
\end{center}

\begin{figure}[!ht]
\centering
\includegraphics[width=\textwidth]{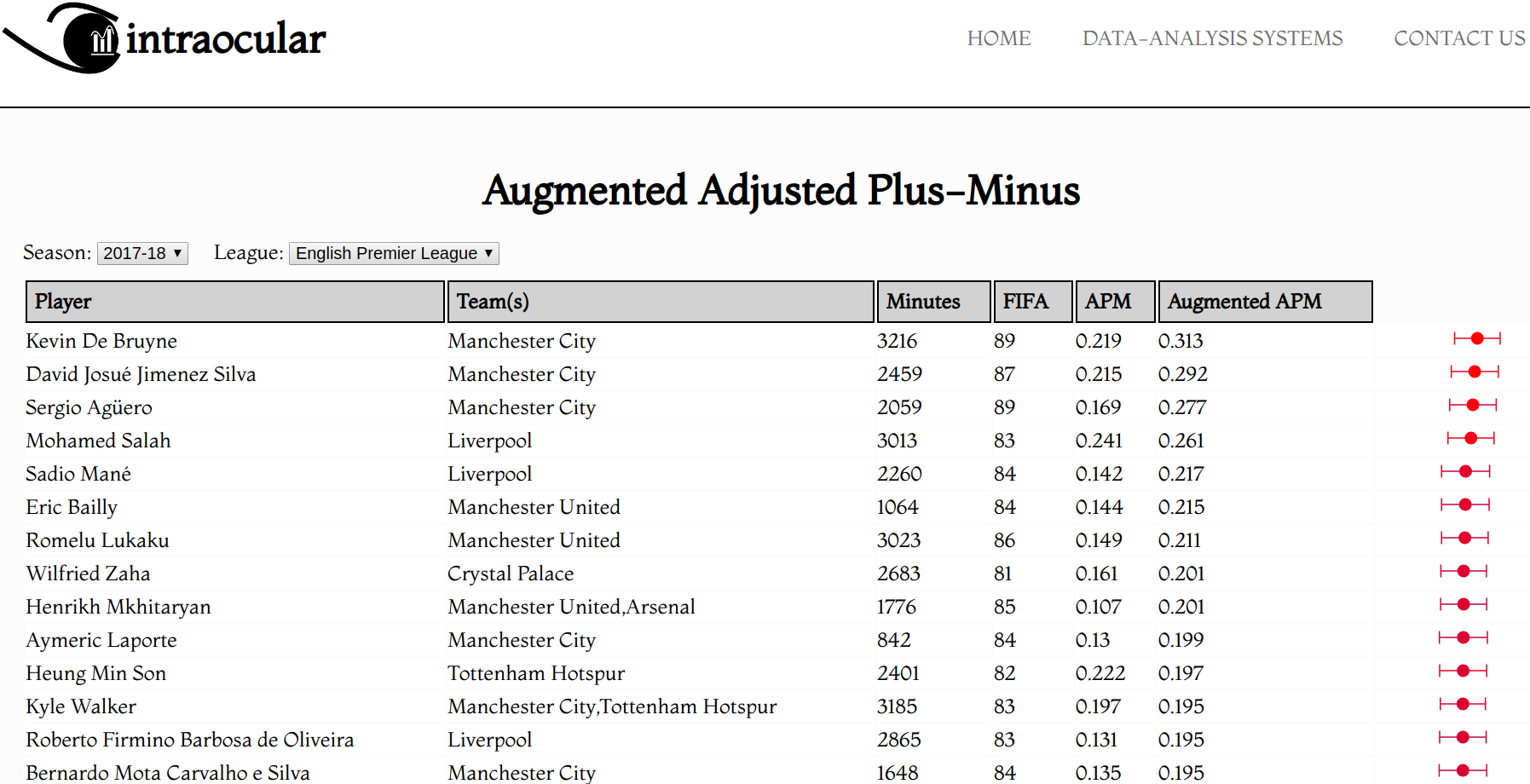}
\caption{The sortable table for our CUDAS, which we make available online, and we continuously update as new games are played. This screen shot displays the top EPL players, sorted by AAPM, for the 2017-18 season.}
\label{fig:intraocular}
\end{figure}

Like ESPN, we made our tables sortable, and an example for the 2017-18 English Premier League season is shown in Figure \ref{fig:intraocular}. We created our sortable tables with the Javascript library {\tt D3} (\cite{bostock2011d3}). 

Finally, we need to make sure our results continuously update. Just as the 2016 election forecast updates after each poll, we want our AAPM statistic to update after each soccer match. I'm not an expert here, but here's three ways you can continuously update results:

\begin{enumerate}
	\item Manually run a script every time you want new results.
	\item Set up a {\tt cron} job (\cite{cronjob}) to run every night.
	\item Use work-flow management tool, such as {\tt Luigi} or {\tt Airflow} (\cite{luigi,airflow}).
\end{enumerate}

Since our AAPM CUDAS is in an early stage, we simply run our scripts manually. But moving forward, we play on switching to an automated workflow. 

And that's how our AAPM CUDAS works. To reiterate, we chose the AAPM statistic because it's {\it interesting} and {\it interpretable}. Then, we built a data pipeline that retrieves data from the web, links together multiple data sources, and produces a design matrix linked to FIFA ratings. We computed AAPM for each player with a Bayesian model, and this provides a ranking of each player in our dataset. Finally, we displayed our results as sortable tables online, and showed how our results can be continuously updated. 

\section{Discussion}
\label{sec:discussion}
I've described a CUDAS as my idealized final product for a data science project. A CUDAS includes a data pipeline, data-analysis, and continuously-updated results, and works for any context. I walked through three examples of successful CUDAS projects, then I described what I thought made them successful. I then described the creation of two CUDAS systems I've been involved with: one for synthetic ecosystems, and one for soccer ratings.

Now that I've explained what a CUDAS is, discussed several examples, and described how they can be built, I want to make the case for thinking about data science projects in terms of a CUDAS. 

The key feature of a CUDAS is that it applies to any context. Let's say you're a statistician who is passionate about the economy. You notice that the GDP statistic is flawed, and you think of a clever way to improve it. What better way to communicate your metric than building a CUDAS?

Now take a more intricate example. In 2018, I gave a presentation where I proposed a CUDAS for my fantasy basketball team (\cite{richardson2018}). In fantasy basketball, you need to decide who to play, but I couldn't find any high quality forecasts tailored to the specifics of my team. So I started making the forecasts myself, but every time I needed an update, I had to manually copy the data from my league into a spreadsheet, manually run the forecast, and only then could I make an informed decision on who to play. This took way too much time, which makes it a perfect opportunity for a CUDAS. In this case, the data pipeline would download my league's data, the data-analysis would prepare the forecasts, and I could display the results in an easy-to-read web page.

Admittedly, building a CUDAS requires skills outside the wheelhouse of statisticians, and we had to pick up a lot of skills along the way. We used the {\tt rvest R} package (\cite{wickham2016package}) for web scraping, we developed an {\it active record linkage} {\tt R} package for name matching, we used the {\it Javascript} library {\it D3} for our interactive tables, we used {\it Google App Engine} to host our website, and so on.

While this took a lot of work, data science tools are quickly maturing, and higher quality tools should enable higher quality CUDAS systems. For example, the {\tt R} package {\tt shiny} has enabled users to create interactive web applications, while requiring zero knowledge of how the web works. And as I mentioned earlier, data pipeline tools have allowed data engineers to streamline and stress-test their data pipelines.

\begin{figure}[!ht]
\centering
\includegraphics[width=\textwidth]{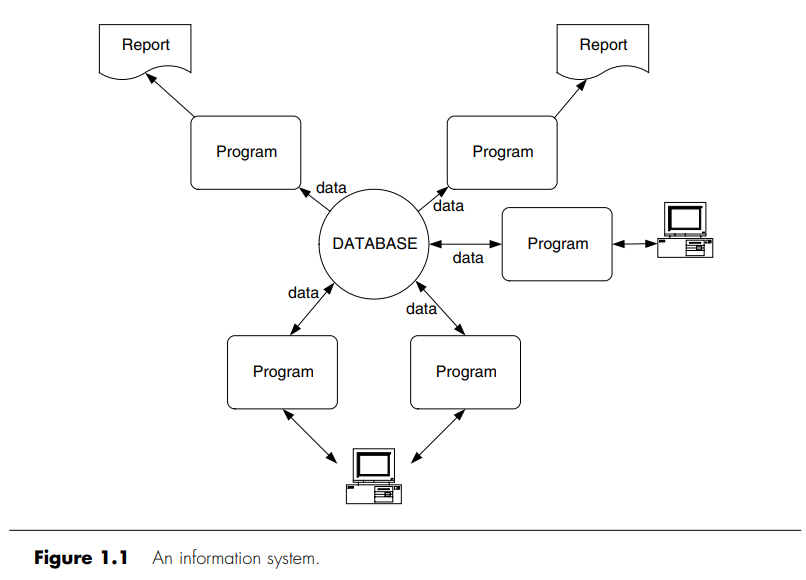}
\caption{An information system as shown in Figure 1.1 of \cite{simsion2004data}.}
\label{fig:informationsystem}
\end{figure}

Further improvements should also come from increased collaboration between data scientists, engineers, computer scientists, web developers, database developers, psychologists, and more. For instance, most of the people I worked with on building our CUDAS systems were statisticians (data scientists?). Our backgrounds were great for developing the statistical models, but our skills were stretched when building the data pipeline, and developing the web applications to display our results. And as I went along, it became clear how valuable data modeling, data visualization, and web development expertise were to building high quality CUDAS systems. I came to see the projects as less about statistical modeling, and more about building an {\it information system} (Figure \ref{fig:informationsystem}). In this way, the CUDAS concept provides a unifying framework for data centered professionals with different skills to rally around. 

As a final point, the rise of the Internet has profoundly changed the way we consume information. This has led to {\it echo chambers}, which Wikipedia describes as:

\begin{center}
``{\it a metaphorical description of a situation in which beliefs are amplified or reinforced by communication and repetition inside a closed system.}'' 
\end{center}

In echo chambers, people lose access to a common set of facts, which makes communication difficult. Can a CUDAS help?

One hypothesis is that high quality CUDAS systems could {\it constrain} the public discourse around a common set of facts. If we all agreed that we want unemployment to be low, and GDP to be high, then we could build a CUDAS to track {\it how well} we're doing. 

Would this work? As a thought experiment, consider the effect of {\it FiveThirtyEight}'s CUDAS on Donald Trump's approval rating (\cite{trumpapproval}). This shows that Trump's approval has ranged between 36.4\% to 47.8\% over the course of his presidency. Now think, when is the last time you heard someone claim that Trump's approval rating is either at 20\% or 80 \%? And how much did this happen in the past? Viewed this way, a CUDAS is analogous to a {\it scoreboard}, since it provides political junkies of all stripes a way to monitor a common set of facts. In spirit, CUDAS systems could complement the ideas of \cite{tetlock2016superforecasting}, who advocate for score keeping of forecasts in the public square.

One of my favorite parts of Donoho's {\it 50 Years of Data Science} is the quote given by \cite{cleveland2001data}:

\begin{center}
{\it \ldots [results in] data science should be judged by the extent to which they enable the analyst to learn from data}. 
\end{center}

It's a great quote. But what if we replaced {\it data analyst} with {\it CUDAS}:

\begin{center}
{\it \ldots [results in] data science should be judged by the extent to which improve a CUDAS}. 
\end{center}

I think this quote works just as well. To me, it's hard to think of any data science research that wouldn't, indirectly or directly, demonstrate it's utility in a CUDAS.

\section*{Acknowledgments}
Thanks to Francesca Matano and Taylor Pospisil for collaborating to build a CUDAS for augmented adjusted plus-minus.  Thanks to Shannon Gallagher, Sam Ventura, Bill Eddy, Jeremy Espino, Shawn Brown, Jay Depasse, and everyone at the Pittsburgh Supercomputer Center who helped in creating SPEW synthetic ecosystems. Thanks to the sports reading and research group, in particular Sam Ventura and Ron Yurko, for their encouragement to initially present the concept.

\bibliographystyle{apa}
\bibliography{references}

\end{document}